\documentclass{article}
\usepackage[english]{babel} 

\usepackage{graphicx}

\usepackage{color}

\begin{document}

\newcommand{\rum}{\rule{0.5pt}{0pt}}
\newcommand{\rub}{\rule{1pt}{0pt}}
\newcommand{\rim}{\rule{0.3pt}{0pt}}
\newcommand{\numtimes}{\mbox{\raisebox{1.5pt}{${\scriptscriptstyle \rum\times}$}}}
\newcommand{\numtimess}{\mbox{\raisebox{1.0pt}{${\scriptscriptstyle \rum\times}$}}}
\newcommand{\Boldsq}{\vbox{\hrule height 0.7pt
\hbox{\vrule width 0.7pt \phantom{\footnotesize T}%
\vrule width 0.7pt}\hrule height 0.7pt}}
\newcommand{\two}{$\raise.5ex\hbox{$\scriptstyle 1$}\kern-.1em/
\kern-.15em\lower.25ex\hbox{$\scriptstyle 2$}$}

\renewcommand{\refname}{References}
\renewcommand{\tablename}{\small Table}
\renewcommand{\figurename}{\small Fig.}
\renewcommand{\contentsname}{Contents}

\begin{center}
{\Large\bf 
Dynamical 3-Space:  Emergent Gravity \rule{0pt}{13pt}}\par

\bigskip
Reginald T. Cahill \\ 
{\small\it  School of Chemical and Physical  Sciences, Flinders University,
Adelaide 5001, Australia}\\
\raisebox{+1pt}{\footnotesize E-mail: Reg.Cahill@flinders.edu.au}\\  
\vspace{3mm}
Invited contribution to \\
\vspace{2mm}
{\it Should the Laws of Gravitation be Reconsidered?\\	
\vspace{-2mm}H\'{e}ctor A. M\'{u}nera, ed. (Montreal: Apeiron 2011)\rule{0pt}{15pt}}
\par

\bigskip

{\small\parbox{11cm}{%
The laws of gravitation devised by Newton, and by Hilbert and  Einstein, have failed many experimental and observational tests, namely the bore hole $g$ anomaly,  flat rotation curves for spiral galaxies,  supermassive black hole mass spectrum, uniformly expanding universe, cosmic filaments, laboratory $G$ measurements,   galactic  EM bending, precocious galaxy formation,..  The response has been the introduction of the new {\it epicycles}: ``dark matter", ``dark energy", and others.  To understand gravity we must restart with the experimental discoveries by Galileo, and following a heuristic argument we are led to a uniquely determined  theory of a dynamical 3-space.   That 3-space exists has been missed from the beginning of physics, although it was 1st directly detected by Michelson and Morley in 1887.  Uniquely generalising the quantum theory to include this dynamical 3-space we deduce the response of quantum matter and show that it results in a new account of gravity, and explains the above anomalies and others.  The dynamical theory for this 3-space involves $G$, which determines the dissipation rate of space by matter,  and $\alpha$, which experiments and observation reveal to be the fine structure constant.  For the 1st time we have a comprehensive account of space and matter and their interaction - gravity.  
\rule[0pt]{0pt}{0pt}}}\medskip
\end{center}

\setcounter{section}{0}
\setcounter{equation}{0}
\setcounter{figure}{0}
\setcounter{table}{0}

\markboth{Cahill R.T.  Dynamical 3-Space:  Emergent Gravity}{\thepage}
\markright{ Cahill R.T.  Dynamical 3-Space:  Emergent Gravity}

\section{Space and Gravity: Back to Galileo}  
Although probably apocryphal Galileo's Learning Tower of Pisa  experiment, showing that objects of different mass have the same free-fall acceleration,  was the first key experimental evidence about the nature of space and gravity.  Galileo  actually  did other experiments that demonstrated that effect.  However, starting with that observation, and building on Kepler's planetary discoveries,  Newton went in a  direction that we now know to be flawed, and which subsequently flawed the generalisation by Hilbert and Einstein.  After some 400 years there is now a futile search for ``dark matter" and ``dark energy" - the epicycle fix-ups of  these flawed theories.  Newton's approach was to assume that Galileo's observations could be explained by assuming that the magnitude of a gravitational force acting on an object with  inertial  mass $m$,  was proportional to the  value $m$, in which case $m$ also acted as a gravitational  mass or charge.  This entailed an equality 
of the inertial mass  and the gravitational mass, which became known as the Weak Equivalence Principle. However, starting from Galileo's observations we can follow a different development, and one based on the following: that the equal gravitational acceleration  of objects with different  masses was caused by the flow of space, which had that acceleration at the location of the masses, and that low-mass matter 
acted as a probe of the space acceleration.  This entails the idea that space exists, is dynamical and directly detectable, but that the velocity of space does not directly affect matter, only its constituent acceleration.  The derivation of the reaction of matter to  the accelerating space had to await the development of the quantum theory of matter, and we find then that gravity is a refraction of the quantum waves, and  is thus an emergent phenomenon. We also briefly show that this account of gravity resolves the above anomalies, and leads to new experimental phenomena and tests.   We also discover that the dynamics of space has two parameters: (i) $G$ describing the dissipative flow of space into matter, and which, for the case of the earth,  has been directly detected by means of spacecraft  earth-flyby Doppler shift data, and (ii) $\alpha\approx 1/137$ - the fine structure constant, which determines a self-interaction coupling constant of the dynamical space, and which bore hole $g$ and black hole mass data reveals to be the fine structure constant.  So the new theory of space and gravity not only provides a well tested theory, but also points to a new unification of space, gravity and the quantum theory. It was pointed out in \cite{Book,Review} that this unification appears to arise from an information theoretic approach to comprehending reality, leading to a quantum foam description of space.  The new theory also explains various so-called relativistic effects, but in a way that does not involve ``spacetime". Indeed the putative predictions of the ``spacetime" formalism are  falsified by experiments. Experiments confirm  instead Lorentz's account of relativistic effects, as being caused by the absolute motion of objects wrt space, and for which the maximum speed is $c$. Experiments show that the speed of light, in vacuum, is anisotropic for an observer moving through space, as 1st detected by Michelson and Morley in 1887, and that the flowing space affects both quantum matter and electromagnetic waves, via its time dependence and/or its speed  inhomogeneity.  The dynamical space also exhibits wave/turbulence effects, usually called ``gravitational waves", and again 1st detected in this experiment. We emphasise that the dynamical space is not a hydrodynamical theory, with some entity flowing through a non-dynamical geometrical space. 

\section{Dynamics of Space}
We begin the heuristic derivation of the dynamics of space, and the emergence of gravity as a quantum matter effect, by assuming that Galileo's  observations suggest the existence of a dynamical space, whose acceleration will be shown to determine the same acceleration of matter, and whose velocity determines the observed anisotropy of the speed of light, with the acceleration determining light bending and gravity as refraction effects.  Physics must employ a covariance formulation, in the sense that ultimately predictions are independent of observers, and that there must also be a relativity principle that relates observations by different observers.  We assume then that space has a structure whose movement, wrt an observer, is described by a velocity field,  ${\bf v}({\bf r}, t)$, at the classical physics level,   at a location ${\bf r}$ and time $t$, as defined by the observer.   In particular the space coordinates ${\bf r}$ define an embedding space, which herein we take to be Euclidean.  At a deeper level space is probably a fractal  quantum foam, which is only approximately embeddable in a 3-dimensional space at a coarse-grained level. This embedding space has no ontological existence - it is not real.  Ironically Newton took this space to be real but unobservable, and so a different concept, and so excluding the possibility that gravity was  caused by an accelerating space.   It is assumed that different observers, in relative uniform motion, relate their description of the velocity field by means of the Galilean Relativity Transformation for positions and velocities.  It is usually argued that the Galilean  Relativity Transformations were made redundant and in error by the Special Relativity Transformations. However this is not so - there exist an exact linear mapping between Galilean Relativity and Special Relativity (SR), differing only by definitions of space and time coordinates \cite{CahillMink}. This implies that the so-called SR relativistic effects are not actual dynamical effects - they are purely artifacts of a peculiar choice of space and time coordinates. In particular Lorentz symmetry is merely a consequence of this  choice of space and time coordinates, and is equivalent to Galilean symmetry \cite{CahillMink}.  Nevertheless Lorentz symmetry remains valid, even though a local preferred frame of reference exists.  Lorentz Relativity, however,  goes beyond Galilean Relativity in that  the limiting speed of systems wrt to the local space causes various so-called relativist effects, such as length contractions and clock dilations.  

The Euler covariant constituent acceleration of space  is then defined by
\begin{equation}
{\bf a}({\bf r},t)= lim_{\Delta t \rightarrow 0}\frac{{\bf v}({\bf r}+{\bf v}({\bf r},t)\Delta t,t+\Delta t) - {\bf v}({\bf r},t)}{\Delta t}  = \frac{\partial {\bf v} }{\partial t}+ ({\bf v}\!\cdot\!{\bf \nabla}){\bf v}\nonumber
\label{eqn:Euler}\end{equation}
which describes  the acceleration of a constituent  by tracking its change in velocity. This means that space has a (quantum) structure that permits its velocity to be defined and detected, which experimentally has been done.  We assume here that the flow has zero vorticity $\nabla \times {\bf v}={\bf 0}$, and then the flow is determined by a scalar function ${\bf v}=\nabla u$. We then need one scalar equation to determine the space dynamics, which we construct by forming the divergence of ${\bf a}$. The inhomogeneous term then determines a dissipative flow caused by  matter, expressed as a matter density, and where the coefficient turns out to be Newton's gravitational constant,
\begin{equation}
\nabla\! \cdot\!\left(\frac{\partial {\bf v} }{\partial t}+ ({\bf v}\!\cdot\!{\bf \nabla}){\bf v}\right) =-4\pi G\rho({\bf r},t) \nonumber
\label{flow}\end{equation}
Note that even a time independent matter density can be associated with a time-dependent flow. This equation follows essentially from covariance and dimensional analysis.  For a spherically symmetric matter distribution, of total mass $M$, and a time-independent spherically symmetric flow we obtain from the above, and external to the sphere of matter, the acceleration of space
$$
{\bf v}({\bf r})=-\sqrt{\frac{2GM}{r}\hat{\bf r}} , \mbox{  giving \ \ } {\bf a}({\bf r})=-\frac{GM}{r^2}\hat{\bf r}
$$
which is an inverse square law.  Newton applied such an acceleration to matter, not space,  and which Newton invented directly by examining Kepler's planetary motion laws, but which makes no mention of what is causing the acceleration of matter, although in a letter in 1675 to Oldenburg, Secretary of the Royal Society, and later to Robert Boyle, he speculated that an undetectable ether flow through space may be responsible for gravity.    Here, however,  the inverse square law emerges from the Euler constituent acceleration, which imposes a space self-interaction. If the sphere of matter is in motion, asymptotically wrt  space, then the flow equation becomes non-trivial to solve, and no analytic solutions are known. Numerical solutions reveal non-trivial wave effects.  Note that one cannot go from a flow of space associated with, say matter asymptotically stationary wrt to space, to the case where the matter is moving, asymptotically, wrt to space - these are very different dynamical situations.  But in either case it is trivial to transform the velocity field, using Galilean Relativity,  between different observers  who are in relative motion.  

While the above 3-space dynamical equation followed from covariance and dimensional analysis, this derivation is not complete yet.  One can add additional terms with the same order in speed and spatial derivatives, and  which cannot be {\it a priori} neglected.  There are two such terms, as in
\begin{equation}
\nabla\! \cdot\!\left(\frac{\partial {\bf v} }{\partial t}+ ({\bf v}\!\cdot\!{\bf \nabla}){\bf v}\right)+
\frac{\alpha}{8}\left((tr D)^2 -tr(D^2)\right)+... =-4\pi G\rho \nonumber \label{eqn:3space}\end{equation}
where $D_{ij}=\partial v_i/\partial x_j$. However to preserve the inverse square law external to a sphere of matter, when the matter is stationary, asymptotically, wrt space, the two terms must have coefficients $\alpha$ and $-\alpha$, as shown.  Here $\alpha$ is a dimensionless space self-interaction coupling constant. The ellipsis denotes higher order derivative terms with dimensioned coupling constants, which come into play when the flow speed changes rapidly wrt separation.  This then gives us the dynamical theory of 3-space. It can be thought of as  arising  via  a derivative expansion from a deeper theory, such as a quantum foam theory \cite{Book}. Note that the equation does not involve $c$, is non-linear and time-dependent, and involves non-local direct interactions. Is success implies that the universe is more connected than previously thought.  Even in the absence of matter there can be time-dependent flows of space. To test this theory we need to determine how quantum matter and EM radiation respond to this dynamical space.  We note immediately that this dynamics is very  rich in that various new phenomena emerge, and which have been observed, and which do not occur in Newtonian gravity, which is a linear theory,  nor in its relativistic generalisation, General Relativity (GR), with both being   one-parameter theories, $G$: essentially GR is flawed by the assumption that GR must reduce to Newtonian gravity in the non-relativistic low-mass limit.

\section{Quantum Matter and Emergent Gravity}
We now derive, uniquely, how quantum matter responds to the dynamical 3-space. This gives the 1st derivation of the phenomenon of gravity, and reveals  this  to be a quantum matter wave refraction effect.  For a free-fall quantum system with mass
$m$ the Schr\"{o}dinger equation is  uniquely generalised   \cite{Schrod}, with the new terms required to maintain that the motion is intrinsically wrt the 3-space, and not wrt the embedding space,  and that the time evolution is unitary
\begin{equation}
i\hbar\frac{\partial \psi({\bf r},t)}{\partial t}  =-\frac{\hbar^2}{2m}\nabla^2\psi({\bf r},t)-i\hbar\left({\bf
v}.\nabla+\frac{1}{2}\nabla.{\bf v}\right) \psi({\bf r},t). \nonumber
\label{eqn:Schrod}\end{equation}
The space and time coordinates $\{t,x,y,z\}$  ensure that  the separation of a deeper and unified process into different classes of phenomena - here a dynamical 3-space (quantum foam) and a quantum matter system, is properly tracked and connected. As well the same coordinates may be used by an observer to also track the different phenomena.  
A quantum wave packet propagation analysis gives  the matter acceleration ${\bf g}=d^2\!\!<\!\!{\bf r}\!\!>\!\!/dt^2$  induced by wave refraction to be \cite{Schrod}
\begin{equation}
{\bf g}=\frac{\partial{\bf v}}{\partial t}+({\bf v}.\nabla){\bf v}+
(\nabla\times{\bf v})\times{\bf v}_R+... \nonumber
\label{eqn:acceln}\end{equation}
 \vspace{-4mm}
\begin{equation}
{\bf v}_R({\bf r}_0(t),t) ={\bf v}_0(t) - {\bf v}({\bf r}_0(t),t),  \nonumber
\end{equation}
where ${\bf v}_R$ is the velocity of the wave packet relative to the 3-space, and where ${\bf v}_O$ and ${\bf r}_O$ are the velocity and position relative to the observer. The last term  generates the Lense-Thirring effect as a vorticity driven effect.  In the limit of zero vorticity we obtain that the quantum matter acceleration is the same as the 3-space acceleration:  ${\bf g}={\bf a}$. This confirms that the new physics is in agreement with Galileo's observations that all matter falls with the same acceleration. Using  arcane language this amounts to a derivation of the Weak Equivalence Principle.

Significantly the quantum matter 3-space-induced  `gravitational' acceleration  also follows from maximising the elapsed proper time wrt the quantum matter wave-packet trajectory ${\bf r}_o(t)$, see \cite{Book},
\begin{equation}
\tau=\int dt \sqrt{1-\frac{{\bf v}^2_R({\bf r}_0(t),t)}{c^2}}  \nonumber
\label{eqn:propertime}\end{equation}
which entails that matter has a maximum speed of $c$ wrt to space, and not wrt an observer. This ensures that quantum waves propagating along neighbouring paths are in phase - the condition for a classical trajectory. This  maximisation gives
\begin{equation}
{\bf g}=\displaystyle{\frac{\partial {\bf v}}{\partial t}}+({\bf v}\cdot{\bf \nabla}){\bf
v}+({\bf \nabla}\times{\bf v})\times{\bf v}_R-\frac{{\bf
v}_R}{1-\displaystyle{\frac{{\bf v}_R^2}{c^2}}}
\frac{1}{2}\frac{d}{dt}\left(\frac{{\bf v}_R^2}{c^2}\right)+...\nonumber
\label{eqn:acceleration}\end{equation}
and then taking the limit $v_R/c \rightarrow 0$ we recover the non-relativistic limit, above. This shows that (i) the matter `gravitational' geodesic is a quantum wave refraction effect, with the trajectory determined by a Fermat maximum proper-time  principle, and (ii) that quantum systems undergo a local time dilation effect. The last, relativistic, term generates the planetary precession  effect.   If clocks are forced to travel different trajectories then the above predicts different evolved times when they again meet - this is the Twin Effect, which now has a simple and explicit physical explanation - it is an absolute motion effect, meaning motion wrt space itself. This elapsed proper time expression invokes Lorentzian relativity, that the maximum speed is $c$ wrt to space, and not wrt the observer, as in Einstein SR. The differential proper time has the form
$$c^2d\tau^2=c^2dt^2-(d{\bf r}-{\bf v}({\bf r},t)dt)^2=g_{\mu\nu}dx^\mu dx^\nu$$
which defines an induced metric for a curved spacetime manifold. However this has no ontological significance, and the metric is not determined by GR.

\section{Electromagnetic Radiation and Dynamical Space}
We must   generalise the Maxwell equations so that the electric and magnetic  fields are excitations within the dynamical 3-space, and not of the embedding space.  The minimal form in the absence of charges and currents is 
 \begin{eqnarray}
\displaystyle{ \nabla \times {\bf E}}&=&\displaystyle{-\mu_0\left(\frac{\partial {\bf H}}{\partial t}+{\bf v.\nabla H}\right)},
 \mbox{\ \ \ }\displaystyle{\nabla.{\bf E}={\bf 0}},  \nonumber \\
 \displaystyle{ \nabla \times {\bf H}}&=&\displaystyle{\epsilon_0\left(\frac{\partial {\bf E}}{\partial t}+{\bf v.\nabla E}\right)},
\mbox{\ \ \  }\displaystyle{\nabla.{\bf H}={\bf 0}}\nonumber\label{eqn:E18}\end{eqnarray}
which was first suggested by Hertz in 1890  \cite{Hertz}, but with ${\bf v}$ then being only a constant vector field, and not interpreted as a moving space effect. As easily determined  the speed of EM radiation is now $c=1/\sqrt{\mu_0\epsilon_0}$ with respect to the 3-space, and not wrt an observer in motion through the 3-space. The Michelson-Morley 1887 experiment 1st detected this anisotropy, as have numerous subsequent experiments. 
A time-dependent and/or inhomogeneous  velocity field causes the refraction of EM radiation. This can be computed by using the Fermat least-time approximation - the opposite of that for quantum matter. This ensures that EM waves along neighbouring paths are in phase. Then the EM ray paths  ${\bf r}(t)$ are determined by minimising  the elapsed travel time: 
\begin{equation}
T=\int_{s_i}^{s_f}\frac{ds\displaystyle{{ |} \frac{d{\bf r}}{ds}|}}{|c\hat{{\bf v}}_R(s)+{\bf v}(\bf{r}(s),t(s)|},  \mbox{  with \ \ }{\bf v}_R=  \frac{d{\bf r}}{dt}-{\bf v}(\bf{r}(t),t) \nonumber
\label{eqn:EMtime}\end{equation}
by varying both ${\bf r}(s)$ and $t(s)$, finally giving ${\bf r}(t)$. Here $s$ is an arbitrary path parameter, and $c\hat {\bf v}_R$ is the velocity of the EM radiation wrt the local 3-space, namely $c$.    The denominator is the speed of the EM radiation wrt the observer's Euclidean spatial coordinates.
This equation may also be used to calculate the gravitational lensing by  black holes, filaments and by ordinary matter, using the appropriate  3-space velocity field.  It produces the measured light bending by the sun.

\section{Dispensing with Dark Matter}
Combining the 3-space zero-vorticity dynamics with the quantum matter acceleration, we obtain
\begin{equation}
\nabla\cdot{\bf g}=-4\pi G\rho-4\pi G\rho_{DM}, \mbox{\ \  } \nabla \times {\bf g}={\bf 0}  \nonumber
\label{eqn:NGplus}\end{equation}
where we define
\begin{equation}
\rho_{DM} = \frac{\alpha}{32\pi G}\left((tr D)^2 -tr(D^2)\right).  \nonumber
\label{eqn:darkmatter}\end{equation}
This is Newtonian gravity but with the extra dynamical term, whose strength is given by $\alpha$. The role of this expression is to reveal that if we analyse gravitational phenomena we will usually find that the matter density $\rho$ is insufficient to account for the observed ${\bf g}$. Until recently this failure of Newtonian gravity has been explained away as being caused by some unknown and undetected but real  ``dark matter" density - an epicycle type explanation.  This expression shows that to the contrary it is a dynamical property of 3-space itself. In deference to that  language we call $\rho_{DM}$ the 3-space induced effective dark matter density. From observed galactic EM lensing and galactic  star trajectories  $\rho_{DM}$ may be  determined and compared with the dynamical 3-space dynamics \cite{Book,Review,Paradigm}.

\section{Spatial Inflows into Spherical Matter}
For  the special case of a spherically symmetric flow  into a spherical matter system at rest, we set  ${\bf v}({ \bf r}, t)=-{\bf {\hat r}} v(r,t)$, and  then the 3-space dynamics becomes, with $v^\prime=\partial v/\partial r$,
\begin{equation}
\frac{\partial v^\prime}{\partial t}+v v''+\frac{2}{r} v v' + (v')^2 + \frac{\alpha}{2 r}\left(\frac{v^2}{2r}+v v' \right) =-4 \pi G \rho  \nonumber
\label{eqn:sphericalsym}\end{equation}
For a  matter density $\rho(r)$ and  total mass $M$, with maximum radius $R$,  this has an exact  static solution \cite{Sun}
\begin{equation}
v(r)^2=
     \displaystyle  \frac{2G}{(1-\frac{\alpha}{2})r} \int_0^r 4 \pi s^2 \rho(s) ds \smallskip \\ 
     \displaystyle  + \frac{2G}{(1-\frac{\alpha}{2})r^\frac{\alpha}{2}} \int_r^R 4 \pi s^{1+\frac{\alpha}{2}} \rho(s) ds , \mbox{\ \  } 0 < r \leq R\nonumber\end{equation}

\begin{equation}
v(r)^2=  \nonumber
     \displaystyle   \frac{2GM}{(1-\frac{\alpha}{2})r} , \qquad r > R
\label{eqn:RawGravitySol}
\end{equation}
 where the apparent mass is $M^*=M/(1-\frac{\alpha}{2})\approx M+ \frac{\alpha}{2} M$, and outside the sphere $g=a=GM^*/r^2$, confirming the observed inverse square law.  The apparent mass is larger than the actual amount of  mater, because of the 3-space self-interaction effect. The exterior inflow speed has been detected for the sun and earth  \cite{CahillNASA}.
 At the center we see a $1/r^{\alpha/2}$ inflow-singularity, but whose strength is mandated by the matter density, and is absent when $\rho(r)=0$ everywhere. This is a minimal attractor or  ``black hole", and is present in all matter systems. The term ``black hole" refers to the existence of an event horizon, where the in-flow speed reaches $c$, but otherwise has no connection to the putative ``black holes" of GR.     The sun, as well as the earth,  has only an  induced  ``minimal attractor'", which affects the interior density, temperature and pressure profiles \cite{Sun}. 
 The 3-space dynamics generates a black hole  effective mass   $M_{BH} \approx \frac{\alpha}{2} M$. These induced  black hole ``effective" masses have been detected in numerous  globular clusters and  spherical galaxies and  their predicted effective masses have been confirmed in some 19 such cases.   The interior non-Newtonian singular inflow effect above is also detectable in bore hole experiments. 3-space is dissipated at the singularity  - the flow does not satisfy a continuity equation - i.e. space is not conserved.

\section{Direct Observation of 3-Space Galactic Flow  and Earth and Sun  Inflows }

Numerous direct observations of 3-space involve the detection of light speed anisotropy. These began with the 1887 Michelson-Morley gas-mode interferometer experiment, that gives a solar system galactic speed in excess of 300 km/s, \cite{MMCK,MMC}. These experiments have revealed components of the flow caused by the sun and the earth, as well as the orbital motion of the earth.  The largest effect is the galactic  velocity of the solar system of 486 km/s in the direction RA = $4.3^\circ$,  Dec = $-75^\circ$,  determined from spacecraft earth-flyby Doppler shift data \cite{CahillNASA}, a direction first detected by Miller in his 1925/26 gas-mode Michelson interferometer experiment \cite{Miller}, and which is completely consistent with the Michelson-Morley data.  The observed  flow component into the sun and into the earth confirms that the 3-space flow is responsible for both  gravity and EM anisotropy \cite{CahillNASA}.

\section{Earth Bore Holes Determine $\alpha$}
The value of the parameter $\alpha$ was first determined from earth bore hole $g$-anomaly data, which shows that gravity decreases more slowly down a bore hole than predicted by Newtonian gravity.  Using the new theory of gravity we find the borehole gravity anomaly at radius $r=R+d$ to be

\begin{equation}
\Delta g=g_{NG}(d)\!-\!g(d)=2\pi \alpha G\rho(R) d+O(\alpha^2), \mbox{\ \ \  } d<0
\label{eqn:ganomaly}\end{equation}
The experimental data then reveals $\alpha$ to be the fine structure constant, to within experimental errors \cite{CahillBH2}.

\section{$G$ Measurement Anomalies}
There has been a long history of anomalies in the laboratory measurements  of Newton's gravitational constant $G$. The explanation is that  the gravitational acceleration external to a piece of matter is only given by application of Newton's inverse square law for the case of an isolated spherically symmetric mass, and using an external small test mass.  For other shapes, and with finite size test masses, the $\alpha$-dependent interaction   results in forces that differ from Newtonian gravity at $O(\alpha)$, as observed. This implies that laboratory measurements to determine $G$ will also measure $\alpha$ \cite{Book}.  

\section{Expanding Universe}
The dynamical 3-space theory  has a time dependent  expanding universe  solution  of the Hubble form.  In the absence of matter, 
$v(r,t)=H(t)r$ with $H(t)=1/(1+\alpha/2)t$, giving a scale factor   $a(t)=(t/t_0)^{4/(4+\alpha)}$, predicting essentially a uniform expansion rate.  This gives a parameter  free account of the supernovae magnitude-redshift data \cite{CahillThermal}. That data reveals a uniformly expanding universe. However the Friedmann equations from GR do not have such a uniformly expanding solution, and {\it ad hoc}  ``dark matter" and ``dark energy"  terms  are added to ``save the theory", giving the current standard cosmological model.   Best fitting the $\Omega_\Lambda$ and $\Omega_{DM}$  $\Lambda$CDM composition parameters to the above solution gives  $\Omega_\Lambda=0.73$ and $\Omega_{DM}=0.27$, the same values as determined by fitting the  $\Lambda$CDM to the supernova data. This demonstrates that ``dark matter" and ``dark energy" are epicycles of  GR. Extending that model into the future leads to the spurious claim that the universe will undergo an exponential rate of expansion.  The search for dark mater and dark energy has now become a cause c\'{e}l\`{e}bre in astronomy, and dominates the NRC decadel plan for astronomy.

\section{Primordial Black Holes}
In the absence of matter the dynamical 3-space equation   has black hole solutions of the form 
$v(r) =- \beta/r^{\alpha/4}$, $ \beta$ arbitrary, giving
$ g(r)=-\alpha\beta/4r^{1+\alpha/2}$,
  as observed in spiral galaxies, resulting in flat rotation curves. These black holes produce a $1/r$ gravitational acceleration, and not a $1/r^2$ form as   assumed in  the usual Newtonian-gravity based analysis of such rotation curves, and then requiring the invention of dark matter.

\section{Primordial Filaments}
The 3-space dynamics also has cosmic filament solutions.  $v(r) =- \mu/r^{\alpha/8}$, where $r$ is here the perpendicular distance from the filament, for arbitrary $\mu$. The gravitational acceleration is   long-range and attractive to matter, i.e. ${\bf g}$ is directed inwards towards the filament,
$g(r) = -\alpha\mu^2/8r^{1+\alpha/4}.$
This is for a single infinite-length filament.  It is conjectured that more complex solutions involving a network of filaments and black holes exist, and which explain the observed cosmic web.

\section{Allais and Other Effects}
The emergent theory of gravity has explained many anomalies, and more are expected to  be revealed, as the dynamics is non-linear and already known to produce both induced and primordial black holes, and filaments.  If masses are not spherical then the induced
black holes produce a long range $1/r$ component of the gavitational acceleration that `leaks' outside of the body. As well bodies that separate from a single body, such as say planets separating from a  centrally forming  star in the early stages, or a moon separating from a planet, will go from one to several induced black holes.  This process may induce a filament joining these black holes.  Such filaments would produce unusual local gravitational accelerations when, for example, a pendulum is oscillating near a filament.  Filaments linking different bodies would also strongly interact  when they begin to overlap.  Such novel phenomena may be the process behind the Allais discovery of unusual pendulum precessions during solar eclipses.     As well the induced black holes result in extremely strong gravity at the center of bodies,  for galaxies, stars and even planets \cite{Sun}.  These result in extremely large central pressures and temperatures in the case of stars and planets.   In the case of planets these central effects would result in conditions akin to those of the earliest seconds of the big bang, when nuclear processes were significant, and when matter was formed from the extremely chaotic dynamics of space, but in the case of planets the conditions prevail over billions of years.  This suggests that there are ongoing matter producing conditions at the centers of planets.  Such a process would cause the planet to  exhibit an ongoing and accelerating expansion, which would also be accompanied by excess heat.  There is observational evidence that the earth is undergoing such an expansion, with the strongest evidence being that the continents fit together most accurately only on a sphere half the present radius of the earth \cite{Carey,Maxlow}.

\section{Conclusions} 
Physics failed to discover the existence of a dynamical 3-space until very recently.  This discovery changes all of physics.  The dynamics has been revealed, and  extensive direct and indirect evidence, from laboratory $G$ measuring experiments, to the expansion of the universe, is now explained.  Only some of that evidence has been cited herein.   The nature of the theory suggests that space is a quantum phenomenon, and the occurrence of the fine structure constant in both quantum matter and space phenomena suggest that a new grand unification of, until now, disparate phenomena is emerging.   As well the experimental data shows that it is Lorentzian relativity that explains relativistic effects, as absolute motion effects, and that Newtonian gravity and its successor, General Relativity,   fail as theories of gravity and, for GR, a theory of the universe.

I thank Professor H\'{e}ctor M\'{u}nera for the opportunity to contribute to this timely volume.


\begin{thebibliography}{99}
     


\bibitem{Book} Cahill  R.T. { \it Process Physics: From Information Theory to Quantum Space
 and Matter},  Nova Science Pub., New York, 2005.   
 \bibitem{Review}  Cahill R.T. {\it Dynamical 3-Space: A Review},   in {\it Ether Space-time and Cosmology: New Insights into a Key Physical Medium},   Duffy M. and L\'{e}vy  J., eds.,  {\it Apeiron}, 135-200, 2009.   
 \bibitem{CahillMink}  Cahill R.T.   {\it  Unravelling Lorentz Covariance and the Spacetime Formalism},   {\it Progress in Physics},  {\bf 4}, 19-24, 2008.

\bibitem{Hertz}  Hertz, H.  {\it On the Fundamental Equations of Electro-Magnetics for Bodies in Motion}, { Wiedemann's Ann.}  { 41}, 369,  1962 {\it  Electric Waves, Collection of Scientific Papers,}   {\it Dover Pub., New  York},  1890.     
\bibitem{Schrod} Cahill R.T. {\it  Dynamical  Fractal  3-Space and the Generalised Schr\"{o}dinger  
Equation: Equivalence Principle and  Vorticity Effects},   {\it Progress in Physics},  {\bf 1}, 27-34, 2006.   
\bibitem{MMCK} Cahill R.T. and Kitto K. {\it Michelson-Morley Experiments Revisited}, {\it Apeiron}, {\bf 10}(2),104-117, 2003.
 \bibitem{MMC}  Cahill  R.T. {\it The Michelson and Morley 1887 Experiment
and the Discovery of Absolute Motion},   {\it Progress in Physics},  {\bf 3}, 25-29, 2005.

 \bibitem{Sun} May R.D. and Cahill R.T. {\it Dynamical 3-Space Gravity Theory: Effects on Polytropic Solar Models},  {\it Progress in Physics}, {\bf 1}, 49-54, 2011. 
    \bibitem{CahillBH2}   Cahill R.T. {\it  3-Space Inflow Theory of Gravity: Boreholes, Blackholes and the Fine Structure Constant},   {\it Progress in Physics},  {\bf 2}, 9-16, 2006.


\bibitem{Paradigm}  Cahill R.T. {\it Unravelling the Dark Matter - Dark Energy Paradigm}, {\it Apeiron}, {\bf 16}, No. 3, 323-375,  2009.
\bibitem{CahillNASA} Cahill R.T. {\it Combining NASA/JPL One-Way Optical-Fiber   Light-Speed  Data with  Spacecraft Earth-Flyby Doppler-Shift Data  to Characterise  3-Space Flow},   {\it Progress in Physics},  {\bf 4}, 50-64, 2009.   
\bibitem{Miller}   Miller D.C. {\it Rev. Mod. Phys.},  {\bf 5}, 203-242, 1933.
  
    

    
    \bibitem{CahillThermal} Cahill R.T. {\it  Dynamical 3-Space Predicts Hotter early Universe: Resolves CMB-BBN $^7$Li and $^4$He Abundance Anomalies},   {\it Progress in Physics},  {\bf 1}, 67-71, 2010.   




\bibitem{Filaments} Cahill R.T. {\it  Dynamical 3-Space:  Cosmic Filaments, Sheets and Voids}, 2011.
       

   
        
\bibitem{Carey}  Carey  S.W. {\it Theories of Earth and Universe.  A History of Dogma in the Earth Sciences}, Stanford Univ. Pres.,1989.
\bibitem{Maxlow} Maxlow J. {\it Terra non Firma Earth. Plate Tectonics is a Myth}, Terrella Press, Australia, 2005.


 

     

\end{thebibliography}
\end{document}